# A Morphological, Topological and Mechanical Investigation of Gyroid, Spinodoid and Dual-Lattice Algorithms as Structural Models of Trabecular Bone

Mahtab Vafaeefar[1], Kevin M. Moerman[2], Majid Kavousi[2], Ted J. Vaughan[1]

[1] Biomechanics Research Centre (BioMEC) and Biomedical Engineering, School of Engineering, College of Science and Engineering, University of Galway, Ireland.
[2] Mechanical Engineering, School of Engineering, College of Science and Engineering, University of Galway, Ireland.

*Address for correspondence:*
Dr Ted J Vaughan
Senior Lecturer in Biomedical Engineering,
Biomechanics Research Centre (BMEC)
Biomedical Engineering
University of Galway
Galway
Ireland
Phone: (+353) 91-493084
Email: ted.vaughan@nuigalway.ie

**Keywords**
Spinodoid, Dual-lattice, Gyroid, Trabecular bone, Computational modelling, Trabecular Structure

Abstract

In this study, we evaluate the performance of three algorithms as computational models of trabecular bone architecture, through systematic evaluation of morphometric, topological, and mechanical properties. Here, we consider the widely-used gyroid lattice structure, the recently-developed spinodoid structure and a structure similar to Voronoi lattices introduced here as the dual-lattice. While all computational models were calibrated to recreate the trabecular tissue volume (e.g., BV/TV), it was found that both the gyroid- and spinodoid-based structures showed substantial differences in many other morphometric and topological parameters and, in turn, showed lower effective mechanical properties compared to trabecular bone. The newly-developed dual-lattice structures better captured both morphometric parameters and mechanical properties, despite certain differences being evident their topological configuration compared to trabecular bone. Still, these computational algorithms provide useful platforms to investigate trabecular bone mechanics and for designing biomimetic structures, which could be produced through additive manufacturing for applications that include bone substitutes, scaffolds and porous implants. Furthermore, the software for the creation of the structures has been added to the open source toolbox GIBBON and is therefore freely available to the community.

# 1 Introduction

Trabecular bone consists of an interconnected plate and rod microarchitecture that is optimised for load-bearing capacity. Deterioration of the microarchitecture of trabecular bone is often observed in bone-related diseases such as osteoporosis (Kirby et al., 2020). Although bone quantity is known to be a major contributor to bone strength (Parkinson et al., 2012), a combination of bone structure and tissue properties are responsible for overall trabecular bone mechanics. However, the precise influence of these parameters on fracture behaviour is not well-understood. Trabecular bone quality is well-described by several parameters that describe the architecture in terms of morphology and topology. Morphological parameters reflect the trabecular bone size and spacing (Lespessailles et al., 2006) and include bone volume fraction (BV/TV), trabecular thickness (Tb.Th), and trabecular separation (Tb.Sp). The structural topology parameters, such as connectivity density (Conn.D) and nodal connectivity, quantify how the constituent parts of trabecular bone are interrelated or arranged (Lespessailles et al., 2006). Computational modelling, through finite element analysis (FEA), has been used extensively to investigate relationships between trabecular bone mechanics and microarchitectural features (Kabel et al., 1999; Koria et al., 2020; Liu et al., 2006; Rammohan et al., 2015). Many of these computational models have been based on micro CT-imaging data of realistic bone from either animal or human samples (Engelke et al., 1996; Jin et al., 2020; Kabel et al., 1999; Koria et al., 2020; Liu et al., 2006; Parkinson et al., 2012; Stauber et al., 2014). However, generating these image-based FEA models requires labour-intensive segmentation and meshing reconstruction. To streamline computational modelling, several computational structures have been proposed to represent trabecular bone microarchitecture (Gibson, 2005; Jin et al., 2020; Kim and Al-Hassani, 2002; Makiyama et al., 2002; Silva and Gibson, 1997) to facilitate a better understanding of the effects of the combined and individual microarchitectural features on mechanical behaviour. Many of the early computational models consist of repeating unit cells (Colabella et al., 2017; Gibson, 2005; Kim and Al-Hassani, 2002), which were applied to the study of age-related microarchitectural changes in trabecular bone (Guo and Kim, 2002). While these have provided useful information on trabecular bone mechanics, many of the proposed computational models have been based on highly idealised geometries, such as open-cell foams (Gibson, 2005), hexagonal columnar structure models (Kim and Al-Hassani, 2002), or gyroid structures (derived from the triply periodic minimal surface family) (Rammohan et al., 2015; Rammohan and Tan, 2016; Society et al., 1996). These repeating unit geometries fail to capture the non-uniformity and overall complexity in the trabecular bone microarchitecture. Furthermore, the input parameters controlling these computational structures are very often limited to bone volume fraction (BV/TV), despite the fact that other morphological and topological factors affect trabecular bone mechanics. This limits their applicability when trying to understand trabecular bone mechanics, and also could limit their application in designing biomimetic structures (Kang et al., 2020) produced through additive manufacturing e.g. for applications that include bone substitutes, scaffolds (Rammohan et al., 2015; Rammohan and Tan, 2016; Timercan et al., 2021; Yánez et al., 2016), and porous implants (Deering et al., 2021).

Recently, studies have included greater complexity when using computational approaches to generate biomimetic trabecular bone structures in an effect to better represent the morphological and topological features of the trabecular bone microarchitecture. Early work by Gibson and co-workers has used Voronoi tessellation in two- (Ruiz et al., 2011, 2010; Schaffner et al., 2000; Silva and Gibson, 1997) and three-dimensions (Chao et al., 2021; Kirby et al., 2020) to create irregular cellular solid representations for low-density trabecular bone (Makiyama et al., 2002; Vajjala et al., 2000). While these structures have provided key insight into the mechanics of age-related changes to trabecular bone, in particular exploring the effect of bone loss on monotonic strength, fatigue properties, and crack growth behaviour, many of these computational structures remain limited by using two-dimensional representations of trabecular bone. Furthermore, these have generally not considered microarchitectural features such as surface curvature, anisotropy, and connectivity. More recently, stochastic three-dimensional porous algorithms such as Gaussian random fields have been suggested as a biomimetic structure for trabecular bone (Callens et al., 2021; Kumar et al., 2020; Roberts and Garboczi, 2002; Zheng et al., 2021). In particular, spinodoid structures have been created by spinodal decomposition during the phase separation process and, unlike repeating unit geometries, these structures visually resemble trabecular bone. With the introduction of tunable anisotropy to the spinodal structures, the generating of smooth cellular architectures with desired structural anisotropy has been enabled (Kumar et al., 2020; Mcfadden et al., 1993; Moelans et al., 2008; Yu and Du, 2006). These features have provided more flexibility for the spinodoid structure to tailor its mechanical response and made it an appealing candidate for trabecular bone modelling. However, detailed microarchitectural properties, and the mechanical responses of the spinodoid structures in comparison to the trabecular bone and other computational candidates, have not been presented. In fact, for the vast majority of computational models that have been proposed (Callens et al., 2021; Kumar et al., 2020; Rammohan et al., 2015; Rodríguez-Montaño et al., 2019; Yánez et al., 2016) to represent the trabecular bone microarchitecture, very few have undergone a systematic evaluation of their morphological, topological and mechanical parameters, as well as the evaluation of their statistical and equivalence to trabecular bone.

In this study, we evaluate the performance of three algorithms as computational models of trabecular bone architecture, through systematic evaluation of morphometric, topological, and mechanical properties. Here, we consider the widely-used gyroid lattice structure (Rammohan et al., 2015; Rammohan and Tan, 2016; Society et al., 1996), the recently-developed spinodoid structure (Kumar et al., 2020; Mcfadden et al., 1993; Moelans et al., 2008; Yu and Du, 2006), and a structure similar to Voronoi lattices (Makiyama et al., 2002; Vajjala et al., 2000) introduced here as the dual-lattice. To evaluate these structures, we analysed their main microarchitectural parameters and studied their mechanical response via FEA. To assess these structures compared to trabecular bone, findings are compared to those for porcine trabecular bone samples derived from CT image data.

## 2 Materials and Methods

### 2.1 Computational microstructures generation

The following three algorithms were considered as candidates to represent the trabecular bone microarchitecture (see Section 2.2):

   I. *Gyroid* structures were built from the triply periodic minimal surface (TPMS) family using a periodic, regular unit-cell approach, which creates an isosurface boundary between the solid and void section of the structure.
  II. *Spinodoid* structures are stochastic models that are derived from a phase-separation process using a Gaussian random field, which is generated by superimposing standing waves of fixed wavelength and amplitude, but random direction and phases.
 III. *Dual-Lattice* structures were built from by thickening the edges of the dual of a tetrahedral Delaunay tessellation.

Each algorithm and the structure produced are fundamentally different and therefore provide a wide design space to explore and compare to trabecular bone. As shown in Figure 1, each of the structures visually resembles trabecular bone microarchitecture. The gyroid and spinodoid structures have been previously used to represent trabecular bone in other studies (Callens et al., 2021; Kumar et al., 2020; Rammohan et al., 2015), however, the dual-lattice structure is implemented here and is similar to Voronoi structures proposed previously (Makiyama et al., 2002; Vajjala et al., 2000). All algorithms were implemented within GIBBON (Moerman, 2018), which is an open source MATLAB toolbox that enables geometry creation, pre-, and post-processing of finite element analysis, as well as analyzing some computational microstructures. All algorithms are available open source through the Gibbon toolbox (Moerman, 2018).

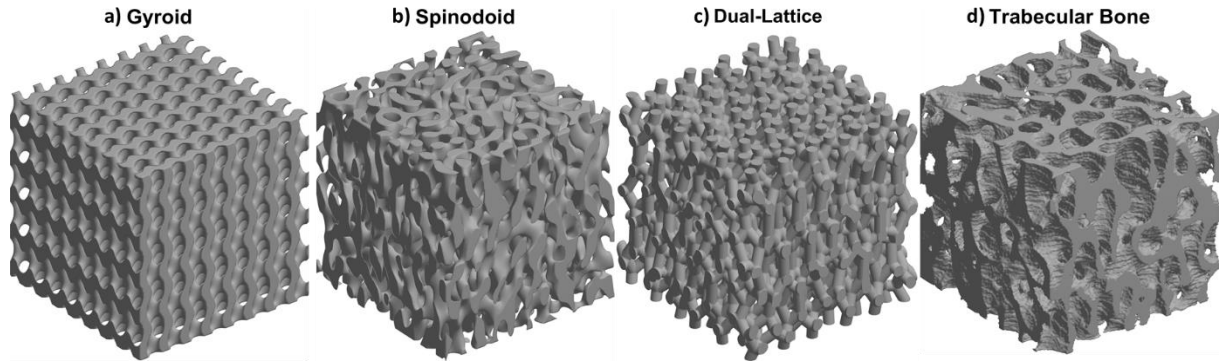

Figure 1. Different structures representing trabecular bone microstructure studied in this paper, a) gyroid structure, b) spinodoid structure, c) dual-lattice structure, and d) real bone CT-image.

### 2.1.1 Gyroid lattice microstructure

The most common types of surface-based unit cells are the TPMS, with a mean curvature of zero (Echeta et al., 2020). The gyroid TPMS structure (Rammohan et al., 2015; Society et al., 1996), investigated in this study is described by the following level-set function $f$ (see also Figure 2(a)):

$$f(x,y,z) = \left[\sin\left(\frac{2\pi}{a_x}x\right)\cos\left(\frac{2\pi}{a_y}y\right)\right]$$
$$+ \left[\sin\left(\frac{2\pi}{a_y}y\right)\cos\left(\frac{2\pi}{a_z}z\right)\right] \quad (1)$$
$$+ \left[\sin\left(\frac{2\pi}{a_z}z\right)\cos\left(\frac{2\pi}{a_x}x\right)\right],$$

where $a_i$, $i = x, y, z$ is the wavelength, representing the periodicity in each direction. Level-set surfaces (see also Figure 2(b)) can be reconstructed using:

$$f(x,y,z) = f_0, \quad (2)$$

where $f_0$ is the level-set parameter. For $f_0 = 0$, the gyroid minimal surface is obtained which subdivides the space into two equal bicontinuous volumes i.e., void and solid (Mangipudi et al., 2016). To achieve other volume fractions, $f_0$ can be altered, e.g., increasing it increases the solid (bone) volume fraction. This allows for the definition of the following binary indicator function $S(x,y,z)$:

$$S(x,y,z) = \begin{cases} 1 \ (bone), & if \quad f(x,y,z) \geq f_0 \\ 0 \ (void), & if \quad f(x,y,z) < f_0 \end{cases} \quad (3)$$

The pure level-set surface (e.g. Figure 2(b)) defines the shell-type gyroid structure (e.g. as used in (Callens et al., 2021)), which is not of interest here. Instead, by closing-over the surface region a lattice-like structure is created (Figure 2(c)). Spatial anisotropy can be applied here by using a different number of periods for each direction, which also imposes mechanical anisotropy. Figure 1(a) shows a resulting gyroid structure of $8 \times 8 \times 5$ unit cells, at the level-set of $f_0 = 0.4$, which resulted in a solid volume fraction of 37%.

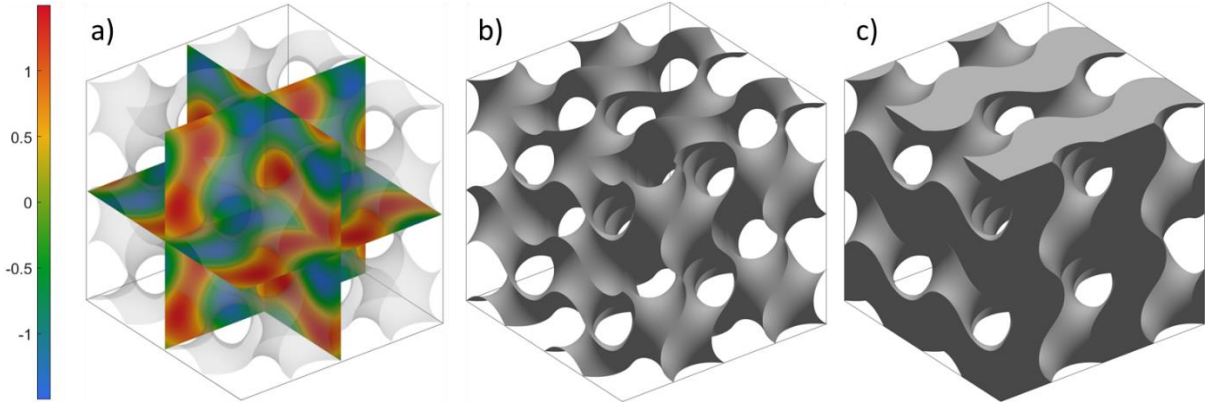

Figure 2. Schematic illustration of the steps to generate the gyroid structures in GIBBON. The gyroid level-set function (a) allows for the construction of the level-set surface (b) which can be closed to form the gyroid structure (c).

### 2.1.2 Spinodoid microstructure

The spinodoid structure is derived from a phase-separation process using a Gaussian random field (GRF) and is generated by superimposing a large number ($N \gg 1$) standing waves of fixed wavelength and amplitude in random directions and phases (Kumar et al., 2020; Soyarslan et al., 2018; Zheng et al., 2021). This GRF function is mathematically described by the following level-set function $f$:

$$f(\mathbf{x}) = \sqrt{\frac{2}{N}} \left[ \sum_{i=1}^{N} \cos(\beta q_i \mathbf{x} + \varphi_i) \right] \quad (4)$$

where $\mathbf{x}$ is the position vector, $\beta$ is the wavenumber, $N$ represents the number of waves considered in the truncated series, and $q_i$ and $\varphi_i$ denote wave direction and wave phase of the $i$th wave, respectively (Soyarslan et al., 2018). The original Cahn-Hilliard equation applies only to isotropic systems, however, by introducing tunable anisotropy factors it results in generating cellular architectures with anisotropic properties (Kumar et al., 2020). This anisotropic extension is implemented by restricting wave directions and biasing their statistical sampling using a non-uniform orientation distribution function (ODF), favouring certain directions, and neglecting others. This is introduced as a direction distribution field whereby,

$$q_i \sim u\left(\{\mathbf{k} \in S^2 : (|\mathbf{k} \cdot \hat{\mathbf{e}}_1| > \cos\theta_1) \oplus (|\mathbf{k} \cdot \hat{\mathbf{e}}_2| > \cos\theta_2) \oplus (|\mathbf{k} \cdot \hat{\mathbf{e}}_3| > \cos\theta_3)\}\right) \quad (5)$$

in which $S^2 = \{\mathbf{k} \in \mathbb{R}^3 : \|\mathbf{k}\| = 1\}$ denoted the units sphere in three dimensional, $\{\hat{\mathbf{e}}_1, \hat{\mathbf{e}}_2, \hat{\mathbf{e}}_3\}$ denotes the Cartesian base vectors and angles $\theta_1, \theta_2, \theta_3 \in \{0\} \cup \left[\theta_{min}, \frac{\pi}{2}\right]$ represent the desired space of wave vectors about each of the three orthogonal base directions. Based on the phase field function, a bi-continuous topology is constructed by computing level-sets of the phase field. For a solid network, this level-set is the quantile evaluated at the average relative density $\rho$ of the solid phase here, and is computed as (Kumar et al., 2020):

$$f_0 = \sqrt{2}\,[\mathrm{erf}^{-1}(2\rho - 1)] \tag{6}$$

where $f_0$ is the level-set parameter. Since small relative densities ($\rho \ll 1$) may contain disjoint solid domains, the design space is limited to $\rho \in [0.3, 1]$ and minimum cone angle as $\theta_{min} = 15°$ (Kumar et al., 2020).

The binary indicator function, $S(x)$ which denotes the presence of solid vs. void at **x** over the domain (Kumar et al., 2020), and be applied both for the solid- and shell-type architectures, is defined by:

$$S(\mathbf{x}) = \begin{cases} 1(bone) & if \quad f(\mathbf{x}) \leq f_0 \\ 0(void) & if \quad f(\mathbf{x}) > f_0 \end{cases} \tag{7}$$

In this topology, we varied several design parameters to manipulate the output structure. These parameters for spinodoid structure were relative density of the solid phase, conical half angles to control anisotropy and the wavenumber. Figure 1(b) shows a representative spinodoid structure, implemented in GIBBON, with 10,000 waves in GRF function, wavenumber of $7\pi$, relative density of 0.37, and conical half angles of $[43°, 43°, 0]$.

### 2.1.3 Dual-lattice microstructure

The dual-lattice derives its name from the fact that the structure is created using the dual of a tetrahedral mesh. The process of obtaining a dual-lattice structure is illustrated in Figure 3. First, TetGen (Si, 2015) is used to create a tetrahedral mesh with desired stretch factor in each direction in a cubic domain (Figure 3(a)), which allows to create anisotropic properties. The dual-lattice skeleton is then generated by connecting the mid-point of faces to the centroid (rather than the circumcircles in Voronoi structures) of tetrahedral elements as in Figure 3(b). We discretised Delaunay triangulation edges with specified spaced nodes on the skeleton model, and for each node, the distance to the surrounding grids is calculated. A contour of distance values for nodes on mid planes are shown in Figure 3(c). These nodes create the strut axis to be thickened with a user's defined magnitude of circulation. As the last step, surrounding surfaces around the strut axis are created and form the final dual-lattice structure, as in Figure 3(d).

Figure 1(c) shows a dual-lattice structure with a strut thickness of 0.16, Delaunay points pacing of 1.97

in a sample size of 3, and an applied stretch factor of [4,4,2] in $xyz$ directions.

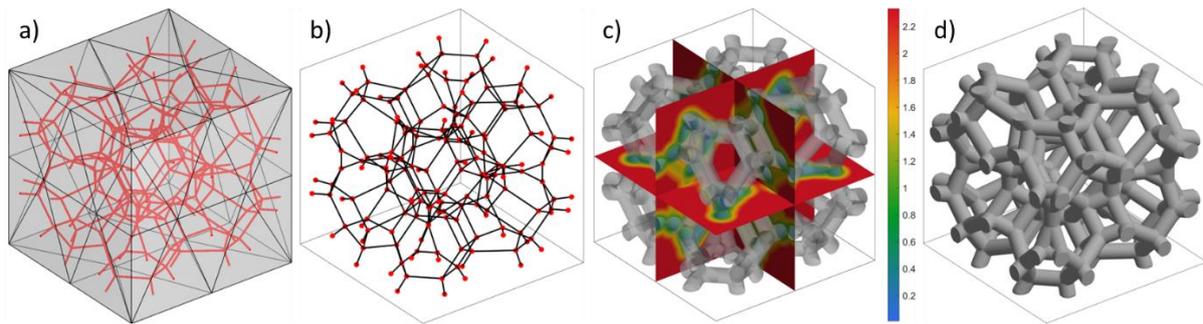

Figure 3. Schematic illustration of the steps to generate the dual-lattice structures in GIBBON. Tetrahedral mesh box (a) and its skeleton generated by connecting mid-point of faces to the centroids of tetrahedra (b). Distance of nodes from neighbouring grids are calculated (c) to create strut thickness with specified distance and forming the final dual-lattice structure after thickening the struts (d).

## 2.2 Trabecular bone data set

Ten micro-CT datasets of porcine talar subchondral trabecular bone samples were obtained from the open access research dataset (Koria et al., 2020). These consisted of cylindrical samples extracted from the talar dome surface having an average gauge length of 6.23 ± 1.29 mm and scanned with an isotropic resolution of 16 µm. From each CT image, several cubic volumes of interest (VOI) of 3 mm³ were extracted and we considered nineteen VOIs in total. The bone samples were segmented using a grayscale threshold in BoneJ (Doube et al., 2010) and 3D binary images of the bone phase were obtained.

## 2.3 Morphometric and Topological characterization

A detailed quantitative morphological and topological characterization of the trabecular bone samples and computationally generated microstructures was conducted. Among the morphological parameters, bone volume fraction (BV/TV), trabecular thickness (Tb.Th), trabecular spacing (Tb.Sp), Structural Model Index (SMI), bone-specific surface area (BS/BV), and structural degree of anisotropy (DA) were analysed. For topological analysis Ellipsoid Factor (EF) (Doube, 2015), connectivity density (Conn.D.) (Odgaard and Gundersen, 1993), Inter-trabecular Angles (ITA) (Reznikov et al., 2016), and nodal connectivity in skeletonized bone network were considered. Table 1 provides a summary of the parameters considered and a brief description of each, most of which were characterized through a combination of MATLAB codes and BoneJ (Doube et al., 2010). The computational models were converted into binary images by using binary indicator functions, defining voxel values as zero and one regarding their position in a void or material section of their domain, respectively. These images were imported to BoneJ as an image stack to be analysed. The trabecular bone images were also binarized and purified prior to connectivity computation in BoneJ since the Euler characteristic is calculated based on a single connected component. While many of these parameters are well-established when analyzing trabecular bone morphometry, and defined in detail elsewhere, here we describe both the ellipsoid factor and the inter-trabecular angle in further detail below.

Table 1. Summary of studied morphometric and topological indices

| Parameter | Abbreviation, unit | Description |
|---|---|---|
| **Bone volume fraction** | BV/TV | The fraction of trabecular bone volume to total tissue volume. |
| **Bone specific surface area** | BS/BV (mm$^{-1}$) | The fraction of trabecular bone surface area to trabecular bone volume. |
| **Trabecular Thickness** | Tb.Th (mm) | The average width of the trabeculae. |
| **Trabecular Separation** | Tb.Sp (mm) | The average distance between trabeculae. |
| **Degree of Anisotropy** | DA | A measure of how highly oriented substructures are within a volume. |
| **Connectivity Density** | Conn.D. (mm$^{-3}$) | The number of closed loops in a unit volume of bone tissue. |
| **Structural Model Index** | SMI | A measure of plate/rod-like geometry of trabecular structures. |
| **Ellipsoid Factor** | EF | A method for measuring rod- and plate-like geometry, independent of other geometric parameters, by fitting maximal inscribed ellipsoids inside 3-dimensional continua (Salmon et al., 2015). |
| **mean Inter-Trabecular Angle** | mean ITA (°) | The mean angle between two connected trabeculae |
| **Number of Neighboring Nodes/ Nodal Connectivity** | $i$-N | The number of trabeculae connected to a junction |

The ellipsoid factor at a point within a 3D structure is defined as the difference in axis ratios of the greatest ellipsoid that fits inside the structure and that contains the point of interest, and ranges from −1 for strongly oblate (discus-shaped) ellipsoids to +1 for strongly prolate (javelin-shaped) ellipsoids (Doube, 2015). In all cases for calculating the ellipsoid factor, the input values were calibrated to provide results with the filling percentage of above 95% of the whole structure and the median change in the ellipsoid factor in the last iteration was below 0.1. Reported values are the mean pixel value of the ellipsoid factor image, as a measure of plate-like or rod-like elements within each structure. Final implementations in this study contain 100 surface points, a sampling increment of 16 μm, a contact sensitivity of 1, and a total of 12 iterations used for each run.

Inter-trabecular angle and nodal connectivity (Reznikov et al., 2016) are topological parameters that are calculated based on a skeletonised 3D plugin in BoneJ (Lee et al., 1994), which applies a thinning algorithm to the trabeculae and represents each trabecula as an edge with exactly one voxel thickness, denoting the local longitudinal axis of each trabeculae (Reznikov et al., 2016). Using AnalyzeSkeleton (Arganda-Carreras et al., 2010), we obtained a set of vectors, defined by the trabeculae start and end point $(x, y, z)$ coordinates. The number of trabeculae connected at a single node gives the nodal connectivity ($i$-Neighbors), a topological property determining the load-bearing veins in a structure and is considered to be a contributor to stiffness (Mangipudi et al., 2016). We analysed the structures' network coordinates to find the nodal connectivity and the inter-trabecular angle for each nodal connectivity cohort of each structure using MATLAB. After skeletonizing the structures in BoneJ, the edges and vertices information were fed to MATLAB, where we analyzed the 3D networks of the structures as a 3D graph. We calculated the nodal connectivity at each of the junctions as well as inter-trabecular angles (ITA) (See Figure S3(a-d) in the Supplementary Document). In all studied microstructures, the number of junctions was between 1100 and 2500.

## 2.4 Finite element simulations

Finite element (FE) simulations were performed using ABAQUS/Standard (v. 6.14, SIMULIA, Dassault Systèmes, 2019). All pre- and post-processing steps were automated and conducted using GIBBON (Moerman, 2018) for all structures considered. Across all structures, the trabecular bone tissue was modelled as isotropic and linear-elastic material with a Young's modulus $E_{tissue}$ = 5 GPa and $v = 0.33$. Uniaxial compressive boundary conditions with displacement along the principal axes were applied to determine responses in the principal directions. In each direction, a total displacement of $u_i = 75$μm ($i = 1,2,3$) (1,2,3 representing x,y and z, respectively) was applied perpendicularly to one face, e.g. $y$, with the other two other faces, $x, z$ restricted in their representative directions as shown in Figure 4(a) on a trabecular bone sample. To approximate uniaxial conditions, a planar constraint was added to all lateral faces. The reaction forces in the three principal directions (the principal trabecular orientation in $z$ and other transverse directions in $x$ and $y$) were analysed for each structure and were used to calculate three orthogonal apparent Young's moduli for each structure. Here, the total reaction force ($RF_i$) was computed as the sum of nodal reaction forces from the prescribed nodes, and the apparent Young's moduli $E_i^*$ by the following formula(Liu et al., 2006):

$$E_i^* = \frac{RF_i \ell_i}{u_i A_i} \tag{8}$$

where $\ell_i$ and $A_i$ were the length and cross-sectional area of the samples in the $i$th direction,

respectively.

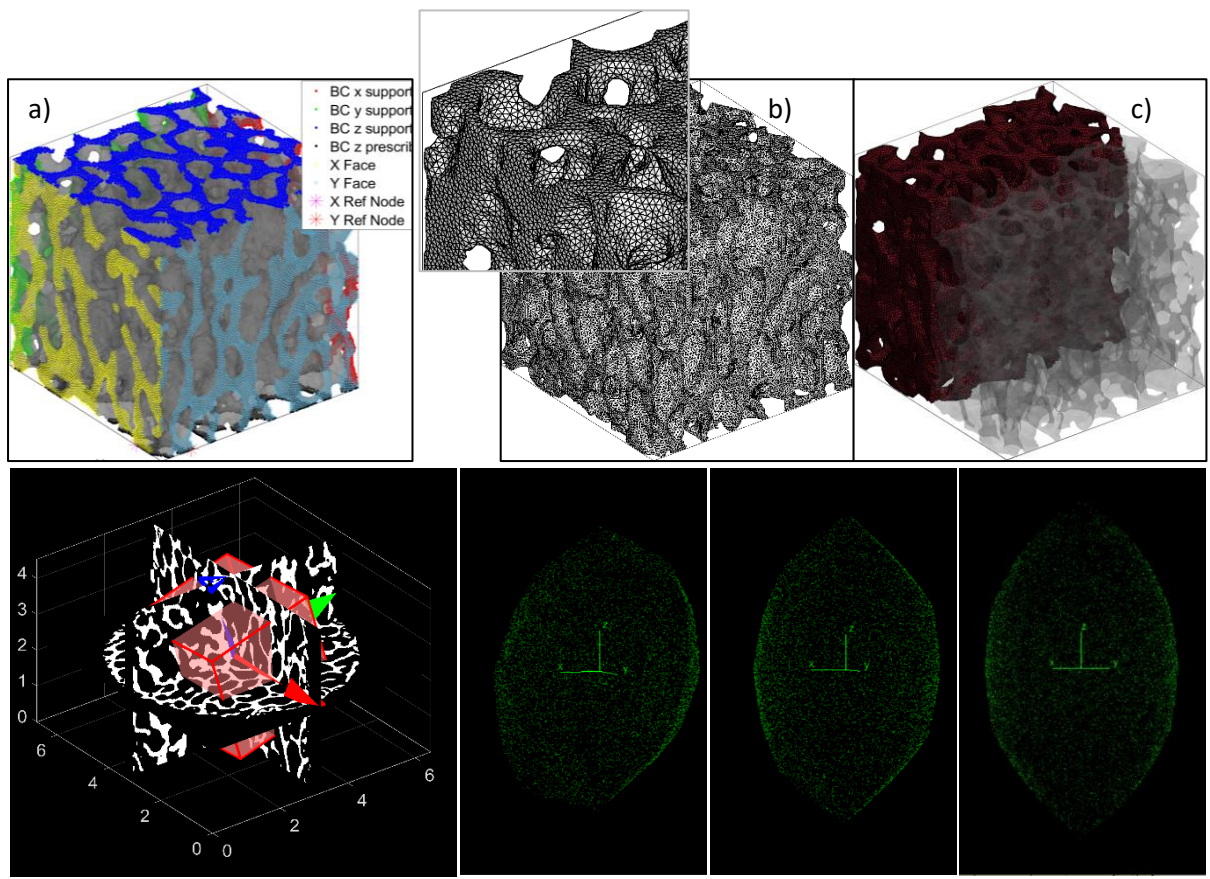

Figure 4. Preparation of CT-image-based finite element models; a) applied boundary condition on the finite element model, b) surface mesh on a trabecular bone sample, c) tetrahedral volume mesh on a trabecular bone sample, d) rotated cubic VOI in principal fabric tensor directions, e) original trabecular bone sample fabric tensor's principal directions, f) cubic bone sample fabric tensor's principal directions, aligned with image principal directions, g) spinodoid structure fabric tensor's principal directions.

### 2.4.1 Surface and volume mesh

Using an isotropic surface meshing algorithm (Lévy and Bonneel, 2013), the surface nodes of the generated geometries were refined with the point spacing to the converged resolution from the image analysis step. The output as the surface mesh was a matrix of triangulated faces that smoothened the curvatures of the original structure. This surface mesh applied to a trabecular bone sample is shown in Figure 4(b). Keeping the largest multiply connected surface and removing the isolated particles, we treated the original surfaces to be fed into our volume meshing algorithm. Volumetric meshing was conducted using TetGen (Si, 2015) toolbox via GIBBON. The final structures were discretized with first-order tetrahedral elements (C3D4). The finite element mesh of a CT-image-based bone sample with a refined tetrahedron mesh network (C3D4) of approximately 1,350,00 elements is shown in Figure 4(c).

Geometrical arrangement and anisotropy of a porous microarchitecture could be characterized by a fabric tensor, which is a symmetric second-order tensor (Cowin, 1985), calculated in BoneJ using the mean intercept length (MIL) method (Odgaard, 1997). The computational models were initially generated with their principal fabric tensors aligned with the image's principal directions. This was confirmed by the ellipsoid's axis and eigenvectors of their fabric tensors shown in Figure 4(g). However, the trabecular bone samples' axis was not aligned with their fabric tensor principal direction. Since the directional stiffness values had to be evaluated on the same geometrical basis, we extracted trabecular bone cubic samples aligned with their fabric tensor principal directions, as shown in Figure 4(d). Figure 4(e) shows the ellipsoid's principal directions of an original trabecular bone sample and Figure 4(f) is the extracted cube sample in fabric tensor principal directions, where its $z$ axis is now aligned with the image $z$ axis.

### 2.5 Determination of RVE size for morphological and mechanical properties

A representative volume element (RVE) is the smallest volume of a heterogeneous material that can be considered statistically representative of the material as a whole, whereby the RVE has the same effective behaviour as the bulk material. To determine the appropriate RVE size for the structures considered here, we conducted a set of simulations considering various RVE sizes and computed the corresponding morphological, topological, and mechanical properties.

To measure the morphometric and topological parameters of the structures, they were converted into images. To make sure that the image-based measurements for determining RVE size were independent of the image's voxel size, and at the same time keeping the computational costs minimum, we conducted a voxel size convergence analysis for our morphological and topological analysis. We considered a cube of $4 \times 4 \times 4$ mm³ of the gyroid, spinodoid, and dual-lattice microstructures at varying voxel sizes, with the highest resolution of 10 μm being considered the reference structure for all cases. Morphological and topological indices were calculated for each resolution, and the relative error of parameters for each resolution was considered against the reference high-resolution values. This analysis is shown in Figure S1(a-e) in the supplementary document, where it was determined that voxel resolutions of 40 μm resulted in less than 10% error in quantifying parameters. All subsequent structures were generated at this resolution.

To determine the appropriate RVE size, eight separate cubic samples, ranging from sizes $1 \times 1 \times 1$ mm³ to $12 \times 12 \times 12$ mm³ were considered, as shown in Figure 6(b), and convergence of morphometric and topological properties were analysed. At each sample size, five separate spinodoid structures were generated, and the relative standard deviation (RSD) was considered as the criteria for spinodoid structure converged RVE size. On the other hand, the gyroid structure and dual-lattice structures only had a single model for each sample size. The error for the biggest sample size of 12 mm was considered as the criteria for determining the suitable RVE in these two structures. The same procedure was applied to their mechanical response. The acceptable RSD in spinodoid structures or error from the biggest sample size in the gyroid and dual-lattice structures is considered to be below 10% for deciding on RVE size in both morphology and mechanical behaviour convergence. It should be noted that some parameters for bigger sample sizes were not analysed due to high computational costs (e.g., dual-lattice structure for sample sizes bigger than 6 mm).

## 2.6 Interrelationship of mechanical and microstructural properties

### 2.6.1 Parameter fit

For capturing the morphological and topological characteristics of trabecular bone samples, we needed to fit the structures to the trabecular bone. Considering the appropriate resolution and RVE size from the previous steps, we manipulated the input parameters of the computational structures to satisfy the bone microarchitecture's morphological and topological parameters. The importance of this calibration originates from the fact that there is not a direct correspondence between input variables and morphological and topological indices. These fitted structures with close-to-bone structural characteristics were considered in FEA to assess their mechanical responses. Trabecular bone-fitted microarchitectures for the gyroid, spinodoid and dual-lattice structure are shown in Figure 1(a-c) besides a CT-image-based cube of $3 \times 3 \times 3$ mm$^3$ of trabecular bone sample in Figure 1(d).

### 2.6.2 Correlation to microarchitecture

The mean, standard deviation (StD), and coefficient of variance (CoV, %) of each morphological, topological, and mechanical property were calculated for bone samples and the computational models. To account for the effect of microarchitectural variations and heterogeneity of trabecular bone, we calculated the correlations between mechanical performance, and the studied morphometric and topological indices of trabecular bone and computational models. Correlation between $E_3^*$ and all morphometric indices were measured using the square of Pearson Product-Moment Correlation Coefficient($r^2$). This statistical analysis was implemented using Microsoft Excel.

To include all possible combinations of paired inputs, and to see the effect of each factor as well as keep the number of tests to a minimum for our computational models, we followed the orthogonal arrays in Taguchi design (Fraley et al., 2007). In our cases, we had two input parameters for the gyroid structure, and three for the spinodoid and dual-lattice structure, respectively. For each input parameter, we considered four levels based on the parameter limits in real trabecular bone samples. As a result, the appropriate orthogonal array was $L16(4^5)$. This means that we needed to have 16 different test cases in each structure with an orthogonal array combination of parameter values, which is shown in Table 2.

Table 2. Orthogonal array design L16 for parameter combinations of gyroid structure and spinodoid structure test cases

| Test ID | Gyroid structure | | Spinodoid structure | | | Dual-lattice structure | | |
|---|---|---|---|---|---|---|---|---|
| | LevelSet | Periods no. | relD[a] | $q_0$[b] | $\Theta$[c](°) | StrutTh[d] | pointSp[e] | Periods no. |
| 1 | 0.64 | [7,7,4] | 0.29 | $6.4\pi$ | [40,40,0] | 0.15 | 1.93 | [4,4,2] |
| 2 | 0.64 | [7,7,5] | 0.29 | $6.6\pi$ | [43,43,0] | 0.15 | 1.95 | [4,4,3] |
| 3 | 0.64 | [8,8,4] | 0.29 | $6.8\pi$ | [48,48,0] | 0.15 | 1.97 | [5,5,2] |
| 4 | 0.64 | [8,8,5] | 0.29 | $7\pi$ | [50,50,0] | 0.15 | 2.00 | [5,5,3] |
| 5 | 0.46 | [7,7,4] | 0.35 | $6.4\pi$ | [43,43,0] | 0.16 | 1.93 | [4,4,3] |
| 6 | 0.46 | [7,7,5] | 0.35 | $6.6\pi$ | [48,48,0] | 0.16 | 1.95 | [5,5,2] |
| 7 | 0.46 | [8,8,4] | 0.35 | $6.8\pi$ | [50,50,0] | 0.16 | 1.97 | [5,5,3] |
| 8 | 0.46 | [8,8,5] | 0.35 | $7\pi$ | [40,40,0] | 0.16 | 2.00 | [4,4,2] |
| 9 | 0.31 | [7,7,4] | 0.40 | $6.4\pi$ | [48,48,0] | 0.17 | 1.93 | [5,5,2] |
| 10 | 0.31 | [7,7,5] | 0.40 | $6.6\pi$ | [50,50,0] | 0.17 | 1.95 | [5,5,3] |
| 11 | 0.31 | [8,8,4] | 0.40 | $6.8\pi$ | [40,40,0] | 0.17 | 1.97 | [4,4,2] |
| 12 | 0.31 | [8,8,5] | 0.40 | $7\pi$ | [43,43,0] | 0.17 | 2.00 | [4,4,3] |
| 13 | 0.12 | [7,7,4] | 0.46 | $6.4\pi$ | [50,50,0] | 0.18 | 1.93 | [5,5,3] |
| 14 | 0.12 | [7,7,5] | 0.46 | $6.6\pi$ | [40,40,0] | 0.18 | 1.95 | [4,4,2] |
| 15 | 0.12 | [8,8,4] | 0.46 | $6.8\pi$ | [43,43,0] | 0.18 | 1.97 | [4,4,3] |
| 16 | 0.12 | [8,8,5] | 0.46 | $7\pi$ | [48,48,0] | 0.18 | 2.00 | [5,5,2] |

[a] relD: relative Density

[b] $q_0$: GRF wave number

[3] Θ: conical half-angles along *xyz* axis

[c] StrutTh: Strut Thickness

[d] pointSp: Point Spacing

## 3 Results

### 3.1 RVE determination

Figure 5 shows the variations of morphological and topological indices for different sample sizes for all three computational models. Considering morphometric and topological parameters, it was found that the spinodoid structure converged within a 5% variation for sample size ≥ 3 mm. Also, for the gyroid and dual-lattice structures, errors for all sample sizes ≥ 3 mm were lower than 10%. Figure 6 shows the mean and individual values of the predicted effective mechanical properties for groups of spinodoid and individual gyroid and dual-lattice structures in each sample size. The predicted mechanical properties for all three structures stabilise for isotropic sample sizes ≥ 3 mm, with a standard deviation and error of less than 7%.

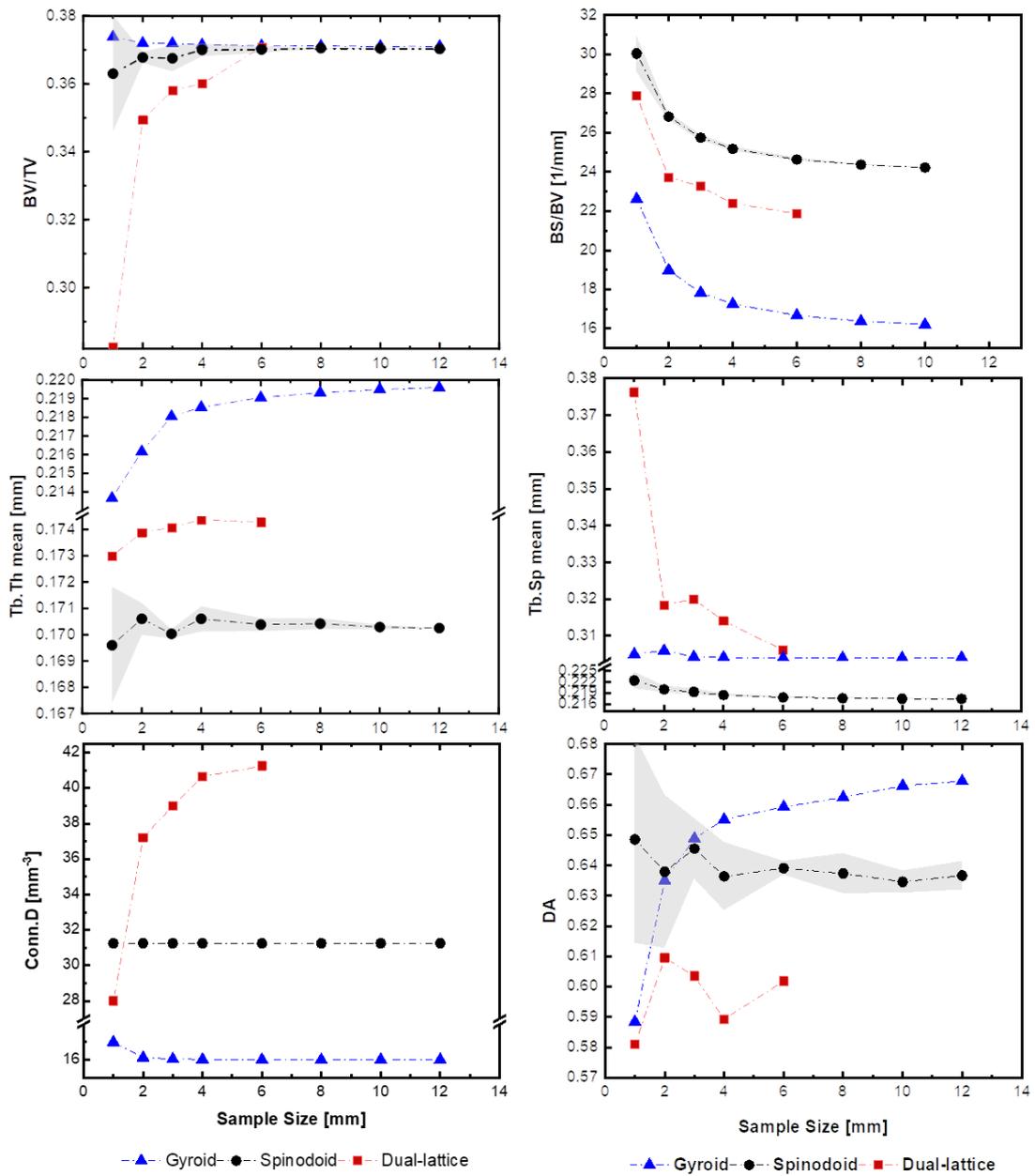

Figure 5. Convergence behaviour of morphometric and topological parameters a) BV/TV, b) BS/BV, c) trabecular thickness, d) trabecular separation, e) connectivity density, f) degree of anisotropy for different sample sizes of all three structures.

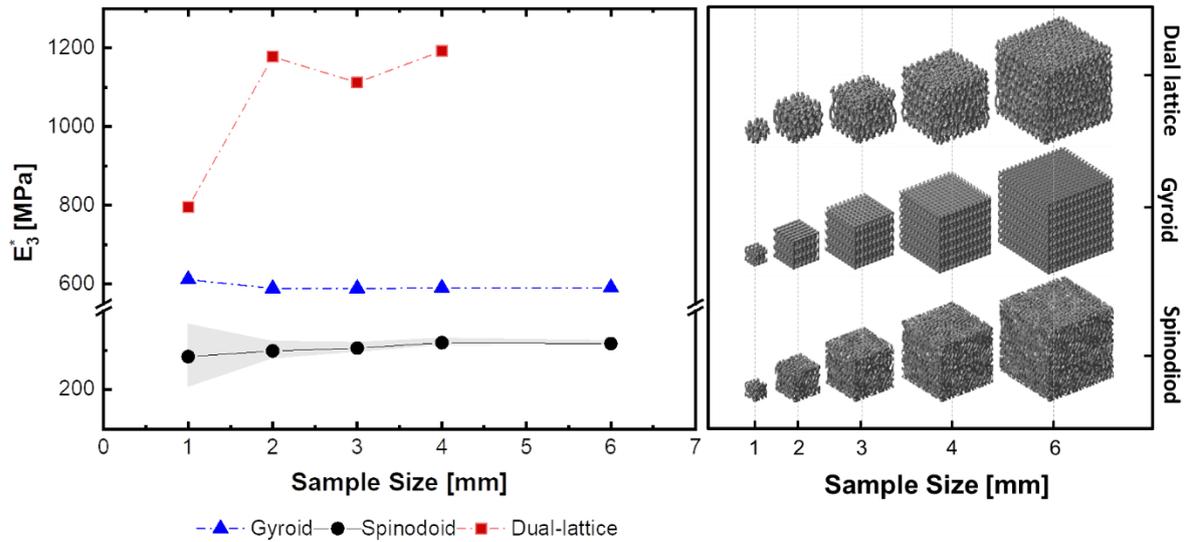

Figure 6. Convergence behaviour of elastic mechanical response for determining RVE size with a resolution of 0.04 mm for gyroid, spinodoid, and dual-lattice structures.

## 3.2 Morphological, topological, and mechanical performance

Considering the mean values of the trabecular bone micro-CT samples as the reference value for each parameter, the input variables were manipulated to generate fitted structures for the trabecular bone. The generated fitted microstructures are shown in Figure 1. A comparison of morphometric, topological, and predicted mechanical properties for all structures, including trabecular bone, is shown in Table 3. Also, Figure 7 shows the 3D colourmap of fitted ellipsoids across all computational models and a trabecular bone sample. Despite many matching morphometric and topological indices, there remain distinct differences between the mechanical responses of computational models, which are likely caused by differences in parameters in each structure.

Table 3. Morphometric and topological parameters and elastic mechanical responses of trabecular bone CT samples, in comparison to those of fitted computational models.

| Parameter | Bone Sample (n=19) | | Gyroid structure | | Spinodoid structure (n=5) | | Dual-Lattice | |
|---|---|---|---|---|---|---|---|---|
| | Mean | Std. dev. | Mean | Std. dev. | Mean | Std. dev. | Mean | Std. dev. |
| BV/TV | 0.36 | 0.04 | 0.37 | n/a | 0.37 | 0.002 | 0.38 | n/a |
| BS/BV (mm$^{-1}$) | 19.83 | 2.15 | 24.06 | n/a | 25.86 | 0.130 | 22.03 | n/a |
| Tb.Th (mm) | 0.16 | 0.02 | 0.17 | n/a | 0.16 | 0.003 | 0.17 | n/a |
| Tb.Sp (mm) | 0.39 | 0.05 | 0.24 | n/a | 0.21 | 0.001 | 0.29 | n/a |
| DA | 0.60 | 0.08 | 0.51 | n/a | 0.64 | 0.007 | 0.62 | n/a |
| Conn.D. (mm$^{-3}$) | 16.55 | 4.50 | 47.44 | n/a | 37.70 | 0.453 | 47.74 | n/a |
| SMI | 0.10 | 0.27 | 1.43 | n/a | 1.77 | 0.022 | 1.32 | n/a |
| EF | -0.17 | 0.02 | -0.06 | n/a | -0.04 | 0.006 | -0.02 | n/a |
| main $i$-N | 3-N | n/a | 3-N | n/a | 3-N | n/a | 4-N | n/a |
| mean ITA (°), 3-N | 117.21 | 0.19 | 120.63 | n/a | 118.26 | 0.124 | n/a | n/a |
| mean ITA (°), 4-N | 108.64 | 0.40 | n/a | n/a | 110.31 | 0.164 | 111.16 | n/a |
| mean ITA (°), 5-N | 104.43 | 0.43 | n/a | n/a | 107.71 | 1.149 | n/a | n/a |
| $E_1^*$(MPa) | 599.41 | 246.61 | 527.79 | n/a | 172.39 | 19.50 | 485.61 | n/a |
| $E_2^*$(MPa) | 740.06 | 182.46 | 526.23 | n/a | 204.50 | 22.90 | 465.33 | n/a |
| $E_3^*$(MPa) | 1187.86 | 222.31 | 886.55 | n/a | 713.51 | 19.94 | 1312.69 | n/a |

The gyroid structure recreates BV/TV and Tb.Th accurately, but it fails to capture several other parameters, like Conn.D., DA, and BS/BV. The gyroid structure under-predicts the mechanical performance in the principal ($E_3^*$) direction compared with the trabecular bone samples. However, in the transverse directions, it recreates mechanical behaviour quite well. Examining both ellipsoid factor and SMI values, the gyroid structure has higher values than the trabecular bone structure, which is indicative of fewer plate-like structures within its architecture. The gyroid is a repeating unit structure, and this results in a repeated distribution pattern of the ellipsoid factor (see Figure 7(a)). Figure 7(b) demonstrates the difference between trabecular bone and computational models visually through a colourmap, confirming that trabecular bone contains more plate-like elements. Also, a visual inspection of the gyroid structure network (Figure S2(a) in supplementary document) shows that the gyroid structure is mainly made up of three connected nodes (3-N).

Since the spinodoid is a stochastic structure, a group of samples containing five members with the same input variables were generated, and the mean values are reported. The variation of morphometric parameters between individual samples is relatively small. From Table 3, the standard deviation in all morphometric and topological indices is below 6%, confirming adequate control over this stochastic microarchitecture. From Table 3, the spinodoid structure captures the three main parameters of BV/TV, Tb.Th and DA accurately. However, other features including Tb.Sp, BS/BV, and SMI parameters deviated from the reference values of the trabecular bone. As a consequence, the spinodoid structure under-predicts the mechanical responses of bone samples in all three principal directions. According to Figure 7(a), spinodoid structures show a higher ellipsoid factor than the trabecular bone. Similar to the gyroid structure, the spinodoid structure also has more rod-like elements rather than plate-like elements. This fact is also confirmed by considering SMI values in spinodoid structure and reference bone. Skeletonised networks show that spinodoid structures, like reference bone samples, have a variety of 3-N to 5-N nodal connectivity, with mean ITA values close to the trabecular bone (Figure S2(b) in supplementary document).

From Table 3, the Dual-lattice structure shows good agreement between the four parameters of BV/TV, Tb.Th, BS/BV and DA when compared to trabecular bone. The Dual-lattice structure has the highest Tb.Sp amongst computational models, which is closer to the trabecular bone values. However, there still remain some limitations as it does not capture Conn.D. or ellipsoid factor accurately. The apparent Young's moduli for the dual-lattice structure were closest to the trabecular bone, both in the principal ($E_3^*$) and transverse ($E_2^*, E_1^*$) directions. Based on Figure 7(e), like in the gyroid structure, the elements are similar and repeated, as a result, we see repeated element distribution for an interval of ellipsoid factor in Figure 7(a) for these two structures. However, as the images of the dual-lattice structures show, it is mainly made up of strut-like or rod-like elements. Being derived from the dual of a tetrahedron mesh, the dual-lattice structure presents solely with 4-N nodal connectivity (Figure S2(c) in supplementary document). Considering that the DA of the dual-lattice structure is equivalent to the trabecular bone, the dramatic increase in $E_3^*$ is likely caused by the 4-N connectivity in the dual-lattice structure instead of the 3-N connectivity in trabecular bone.

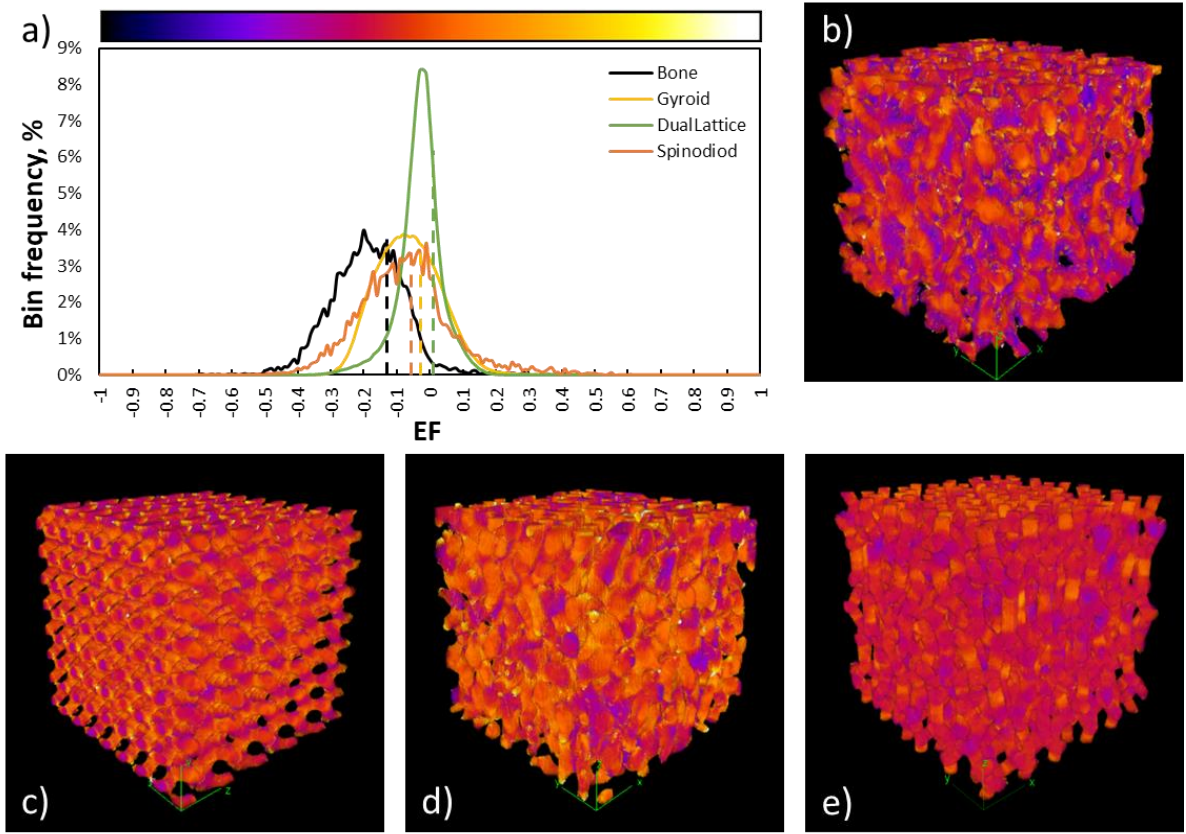

Figure 7. The ellipsoid factor is a measure of plate-like and rod-like elements in structures. a) pixel values of EF histograms for a trabecular bone sample, the gyroid, spinodoid, and dual-lattice microstructures. The dashed vertical lines in between are showing the median EF for each microstructure. b) 3D colour map images of fitted ellipsoids, indicating EF>0 in orange-yellow and EF<0 in purple-blue for the trabecular bone sample, c) gyroid structure, d) spinodoid structure, and e) dual-lattice.

### 3.3  Correlations to microarchitecture

In Figure 8, the correlation of the morphological and topological parameters with the predicted elastic modulus $E_3^*$ is shown for all structures considered. Here, a strong correlation between BV/TV and mechanical response is observed for all three structures, indicating that the computational models are capturing the main microarchitectural correlation in real bone samples. BV/TV, DA, and Tb.Th are positively correlated with an apparent modulus of elasticity in bone and all computational structures, whereas Tb.Sp, BS/BV, SMI factor, and ellipsoid factor are inversely correlated. Conn.D. in trabecular bone and computational models are negatively correlated, except for the spinodoid structures, which unlike the two other computational models and trabecular bone samples, are highly positively correlated to the elastic modulus. According to Figure 8, the correlations of computational models are higher than the trabecular bone samples. A higher correlation of the computational models in comparison to the bone samples could be explained by the computational origin of these structures. Also, there are no direct relationship between these structural parameters and the input variables. Correlations reported for the trabecular bone here are lower than normal values reported in other research (Greenwood et al., 2015; Nazarian et al., 2007), which could be due to the small number of samples used in this study.

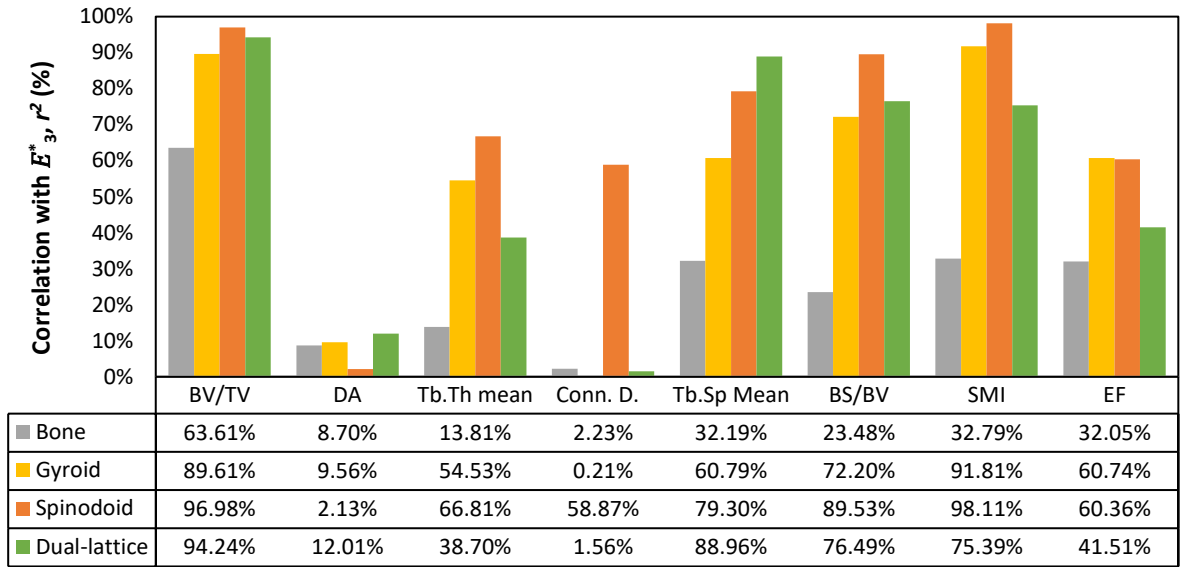

Figure 8. The percentage correlation of morphological and topological indices on variations in the computational model values of apparent Young modulus values as calculated using the correlation coefficient values, $r$ for bone specimens, gyroid, spinodoid and dual-lattice structures.

## 4  Discussion

In this study, we systematically evaluated the morphometric, topological, and mechanical properties of several computational models used to represent trabecular bone structure. In addition to considering gyroid- and spinodoid-based structures, we also developed and implemented a new dual-lattice computational model for trabecular bone based on Delaunay triangulation. While all computational models were calibrated to recreate the trabecular tissue volume (e.g., BV/TV), it was found that both the gyroid- and spinodoid-based structures showed substantial differences in other morphometric and topological parameters and, in turn, showed lower stiffness to the trabecular tissue. The newly proposed dual-lattice structures better captured morphometric parameters and mechanical properties, despite certain differences being evident in its topology compared to trabecular bone e.g., in EF or SMI. These structures are not based on CT images but are computationally affordable to generate and capable of capturing the main morphometric indices. Unlike previous models, these algorithms provide researchers with a large flexible design space to produce a wide range different parameters and capture certain morphometric or topological parameters of trabecular bone.

Until now, many lattice-based models have been proposed (Callens et al., 2021; Colabella et al., 2017; Guo and Kim, 2002; Rammohan and Tan, 2016) but very few have undergone a detailed evaluation of their morphological, topological, and mechanical equivalence to trabecular bone. Despite this, these structures have been used extensively to study trabecular bone mechanics (Colabella et al., 2017; Gibson, 2005; Kim and Al-Hassani, 2002) and the impact of bone loss due to age-related changes and diseases such as osteoporosis. Our study demonstrates that several of these computational models show substantial deviation from the trabecular bone microarchitecture when a broad range of morphometric and topological parameters are considered. For example, the gyroid structure only replicated a limited number of morphometric parameters, with substantial deviation in other morphometric and topological parameters, which meant that the mechanical performance showed a substantial difference from the trabecular bone. Furthermore, the spinodoid structure, which has recently been proposed as a structure that recreates the stochastic nature of trabecular bone (Kumar et al., 2020), also failed to capture most of the structural features and the associated mechanical properties. A key reason for this could be the distribution of plate-like and rod-like elements, which are local classifications of trabecular bone architecture, quantified by the ellipsoid factor and SMI metrics, and are believed to play an important role in the mechanical behaviour of cellular solids (Doube, 2015; Liu et al., 2006; Saha et al., 2010; Stauber and Müller, 2006; Wang et al., 2015). According to the results, the trabecular bone samples have the lowest ellipsoid factors and SMI values among other computational models, meaning that plate-like elements play an important role in the mechanical response of the structure.

Within this study, we introduced the new dual-lattice structure that, besides showing similar morphometric properties as bone, firmly replicates its elastic properties and had the closest stiffness values to trabecular bone, compared to the two other computational models. Despite the dual-lattice structure having a higher ellipsoid factor and SMI value, it appears that the mechanical response of this structure is compensated through increased nodal connectivity in its skeletonized network. While connectivity density is higher in all computational models compared to trabecular bone, the dual-lattice structure has predominantly 4-N connectivity, compared to 3-N connectivity in the trabecular bone network (and other computational models). These differences arise as the connections in the trabecular bone structure are in a more dispersed pattern, rather than being multiply-connected and elongated rod-like units in the computational models. This fact is also depicted in the lower number of junctions in skeletonized model analysis. As more junctions appear throughout the computational models, which is reflected by the higher connectivity density, the trabecular elements form a more compact arrangement, resulting in lower values of trabecular spacing (Tb.Sp). Also, the lower junction density in the trabecular bone samples explains its lower connectivity density compared with the generated models. The more junctions available in a specific volume, the more multiply connected the structure is and though connectivity density increases (Odgaard and Gundersen, 1993). As research suggests (Reznikov et al., 2016), the mean ITA in human trabecular bone approaches the expected "ITA" in a geometrically idealised 3D motif, where all the edges and nodes are equally distant from each other, and maximally span the 3D space. Comparing the measured mean ITA for 3-N and 4-N nodes in bone and computational models, they all follow the idealized distribution of edges to maximally span in 3D space.

Microarchitecture correlation with $E_3^*$ values in Figure 8 show that dual-lattice and gyroid structures are better at predicting trabecular bone behaviour. Unlike the others, the $E_3^*$ values for spinodoid structures are inversely correlated to Conn.D. and over-predicted correlations with most other factors. In general, computational models show higher correlations for $E_3^*$ with the microstructural factors compared to trabecular bone. This could be simply explained by the fact that the structures are computationally dependent on the input variables and are mathematically correlated inherently. However, the variations in trabecular bone originate from the heterogeneity of bone tissue, bone remodelling and different loading conditions which is a mechanobiological response of bone tissue to the changes in its external mechanical environment.

Besides the advantages of these new algorithms compared to previous studies, they still have certain limitations. First of all, there are no one-to-one relationships between the input variables and the measured structural parameters. Therefore, the microarchitectural parameters in these algorithms cannot be changed individually since they are mathematically correlated to each other; e.g. changes in connectivity density will result in changes in elements' thickness or separation. As a result, the user cannot directly control the contribution of individual parameters to the mechanical response of the structure. Also, in this study, we looked at the porcine trabecular bone, not human tissue. However, these methods could be easily applied to human samples, especially when open source codes are available. Finally, local changes have not be applied to these structures; Any changes in input parameters will result in a uniform change in the structures in every element, e.g changing the thickness of individual elements, or losing the connectivity in a specific loop, which happens in osteoporosis, has not been implimented in these models. While these structures do not exactly copy all the topological properties, it was found that they could be tuned to closely match the elastic mechanical properties. Furthermore, all computational models are tunable to provide a degree of anisotropy that is in a similar range to trabecular bone structure. It is also likely that these structures could prove useful in tissue engineering applications, whereby scaffolds can be created through additive manufacturing (Deering et al., 2021). In these applications, the structures presented here could be tailed for mechanical properties and also hydraulic parameters, such as porosity and permeability, which can be of interest to ensure sufficient permeability for cell and tissue in-growth.

## 5 Conclusion

In summary, we considered three computational algorithms and quantitatively analysed their microarchitectures and elastic mechanical behaviour in comparison to a group of trabecular bone samples. The results showed that several commonly used computational models (e.g. the gyroid and spinodoid structures) showed substantial differences in morphometric and topological parameters compared to the trabecular bone samples and, as a result, had lower mechanical properties. We also developed and implemented an entirely new dual-lattice computational model for trabecular bone based on Delaunay triangulation. This dual-lattice structure better captured key morphometric parameters and mechanical properties, despite certain differences being evident in its topology compared to the trabecular bone.

## 6 Acknowledgement

This project has received funding from the European Research Council (ERC) under the EU's Horizon 2020 research and innovation program (Grant agreement No. 804108).

The authors wish to acknowledge the Irish Centre for High-End Computing (ICHEC) for the provision of computational facilities.

# 7 References

v. 6.14, SIMULIA, Dassault Systèmes, 6.14. ed, 2019. . Dassault Systèmes Simulia Corp.

Arganda-Carreras, I., Fernández-González, R., Muñoz-Barrutia, A., Ortiz-De-Solorzano, C., 2010. 3D reconstruction of histological sections: Application to mammary gland tissue. Microsc Res Tech 73, 1019–1029. https://doi.org/10.1002/jemt.20829

Callens, S.J.P., Tourolle né Betts, D.C., Müller, R., Zadpoor, A.A., 2021. The local and global geometry of trabecular bone. Acta Biomater 130, 343–361. https://doi.org/10.1016/j.actbio.2021.06.013

Chao, L., Jiao, C., Liang, H., Xie, D., Shen, L., Liu, Z., 2021. Analysis of Mechanical Properties and Permeability of Trabecular-Like Porous Scaffold by Additive Manufacturing. Front Bioeng Biotechnol 9, 1–13. https://doi.org/10.3389/fbioe.2021.779854

Colabella, L., Cisilino, A.P., Häiat, G., Kowalczyk, P., 2017. Mimetization of the elastic properties of cancellous bone via a parameterized cellular material. Biomech Model Mechanobiol 16, 1485–1502. https://doi.org/10.1007/s10237-017-0901-y

Cowin, S.C., 1985. The relationship between the elasticity tensor and the fabric tensor. Mechanics of Materials 4, 137–147. https://doi.org/10.1016/0167-6636(85)90012-2

Deering, J., Dowling, K.I., DiCecco, L.A., McLean, G.D., Yu, B., Grandfield, K., 2021. Selective Voronoi tessellation as a method to design anisotropic and biomimetic implants. J Mech Behav Biomed Mater 116, 104361. https://doi.org/10.1016/J.JMBBM.2021.104361

Doube, M., 2015. The ellipsoid factor for quantification of rods, plates, and intermediate forms in 3D geometries. Front Endocrinol (Lausanne) 6, 1–5. https://doi.org/10.3389/fendo.2015.00015

Doube, M., Kłosowski, M.M., Arganda-Carreras, I., Cordelières, F.P., Dougherty, R.P., Jackson, J.S., Schmid, B., Hutchinson, J.R., Shefelbine, S.J., 2010. BoneJ: Free and extensible bone image analysis in ImageJ. Bone 47, 1076–1079. https://doi.org/https://doi.org/10.1016/j.bone.2010.08.023

Echeta, I., Feng, X., Dutton, B., Leach, R., Piano, S., 2020. Review of defects in lattice structures manufactured by powder bed fusion. International Journal of Advanced Manufacturing Technology 106, 2649–2668. https://doi.org/10.1007/s00170-019-04753-4

Engelke, K., Song, S.M., Glüer, C.C., Genant, H.K., 1996. A digital model of trabecular bone. Journal of Bone and Mineral Research 11, 480–489. https://doi.org/10.1002/jbmr.5650110409

Fraley, S., Oom, M., Terrien, B., Zalewski, J., Bredeweg, R., Morga, J., Sekol, R., Wong, R., 2007. 4.1:

Design of Experiments via Taguchi Methods-Orthogonal Arrays.

Gibson, L.J., 2005. Biomechanics of cellular solids. J Biomech 38, 377–399. https://doi.org/10.1016/j.jbiomech.2004.09.027

Greenwood, C., Clement, J.G., Dicken, A.J., Evans, J.P.O., Lyburn, I.D., Martin, R.M., Rogers, K.D., Stone, N., Adams, G., Zioupos, P., 2015. The micro-architecture of human cancellous bone from fracture neck of femur patients in relation to the structural integrity and fracture toughness of the tissue. Bone Rep 3, 67–75. https://doi.org/10.1016/j.bonr.2015.10.001

Guo, X.E., Kim, C.H., 2002. Mechanical consequence of trabecular bone loss and its treatment: A three-dimensional model simulation. Bone 30, 404–411. https://doi.org/10.1016/S8756-3282(01)00673-1

Jin, Y., Zhang, T., Cheung, J.P.Y., Wong, T.M., Feng, X., Sun, T., Zu, H., Sze, K.Y., Lu, W.W., 2020. A novel mechanical parameter to quantify the microarchitecture effect on apparent modulus of trabecular bone: A computational analysis of ineffective bone mass. Bone 135, 115314. https://doi.org/10.1016/j.bone.2020.115314

Kabel, J., Odgaard, A., van Rietbergen, B., Huiskes, R., 1999. Connectivity and the elastic properties of cancellous bone. Bone 24, 115–120. https://doi.org/10.1016/S8756-3282(98)00164-1

Kang, J., Dong, E., Li, D., Dong, S., Zhang, C., Wang, L., 2020. Anisotropy characteristics of microstructures for bone substitutes and porous implants with application of additive manufacturing in orthopaedic. Mater Des 191, 108608. https://doi.org/10.1016/j.matdes.2020.108608

Kim, H.S., Al-Hassani, S.T.S., 2002. A morphological model of vertebral trabecular bone. J Biomech 35, 1101–1114. https://doi.org/10.1016/S0021-9290(02)00053-2

Kirby, M., Morshed, A.H., Gomez, J., Xiao, P., Hu, Y., Guo, X.E., Wang, X., 2020. Three-dimensional rendering of trabecular bone microarchitecture using a probabilistic approach. Biomech Model Mechanobiol 19, 1263–1281. https://doi.org/10.1007/s10237-020-01286-8

Koria, L., Mengoni, M., Brockett, C., 2020. Estimating tissue-level properties of porcine talar subchondral bone. J Mech Behav Biomed Mater 110. https://doi.org/10.1016/j.jmbbm.2020.103931

Kumar, S., Tan, S., Zheng, L., Kochmann, D.M., 2020. Inverse-designed spinodoid metamaterials. NPJ Comput Mater 6, 1–10. https://doi.org/10.1038/s41524-020-0341-6

Lee, T.C., Kashyap, R.L., Chu, C.N., 1994. Building Skeleton Models via 3-D Medial Surface Axis Thinning

Algorithms. CVGIP: Graphical Models and Image Processing 56, 462–478. https://doi.org/https://doi.org/10.1006/cgip.1994.1042

Lespessailles, E., Chappard, C., Bonnet, N., Benhamou, C.L., 2006. Imaging techniques for evaluating bone microarchitecture. Joint Bone Spine 73, 254–261. https://doi.org/10.1016/J.JBSPIN.2005.12.002

Lévy, B., Bonneel, N., 2013. Variational Anisotropic Surface Meshing with Voronoi Parallel Linear Enumeration, in: Jiao, X., Weill, J.-C. (Eds.), Proceedings of the 21st International Meshing Roundtable. Springer Berlin Heidelberg, Berlin, Heidelberg, pp. 349–366.

Liu, X.S., Sajda, P., Saha, P.K., Wehrli, F.W., Guo, X.E., 2006. Quantification of the Roles of Trabecular Microarchitecture and Trabecular Type in Determining the Elastic Modulus of Human Trabecular Bone. J Bone Miner Res 21, 1608–1617. https://doi.org/10.1359/JBMR.060716

Makiyama, A.M., Vajjhala, S., Gibson, L.J., 2002. Analysis of crack growth in a 3D voronoi structure: A model for fatigue in low density trabecular bone. J Biomech Eng 124, 512–520. https://doi.org/10.1115/1.1503792

Mangipudi, K.R., Epler, E., Volkert, C.A., 2016. Topology-dependent scaling laws for the stiffness and strength of nanoporous gold. Acta Mater 119, 115–122. https://doi.org/10.1016/j.actamat.2016.08.012

Mcfadden, G.B., Wheeler, A.A., Braun, R. 3, Coriell, S.R., Sekerkat, R.F., 1993. Phase-field models for anisotropic interfaces. PHYSICAL REVIEW E VOLUME 48.

Moelans, N., Blanpain, B., Wollants, P., 2008. An introduction to phase-field modeling of microstructure evolution. CALPHAD 32, 268–294. https://doi.org/10.1016/j.calphad.2007.11.003

Moerman, M.K., 2018. GIBBON: The Geometry and Image-Based Bioengineering add-On. J Open Source Softw 3, 506. https://doi.org/10.21105/joss.00506

Nazarian, A., Muller, J., Zurakowski, D., Müller, R., Snyder, B.D., 2007. Densitometric, morphometric and mechanical distributions in the human proximal femur. J Biomech 40, 2573–2579. https://doi.org/10.1016/j.jbiomech.2006.11.022

Odgaard, A., 1997. Three-dimensional methods for quantification of cancellous bone architecture. Bone 20, 315–328. https://doi.org/https://doi.org/10.1016/S8756-3282(97)00007-0

Odgaard, A., Gundersen, H.J.G., 1993. Quantification of connectivity in cancellous bone, with special emphasis on 3-D reconstructions. Bone 14, 173–182. https://doi.org/10.1016/8756-

3282(93)90245-6

Parkinson, I.H., Badiei, A., Stauber, M., Codrington, J., Müller, R., Fazzalari, N.L., 2012. Vertebral body bone strength: The contribution of individual trabecular element morphology. Osteoporosis International 23, 1957–1965. https://doi.org/10.1007/s00198-011-1832-6

Rammohan, A.V., Lee, T., Tan, V.B.C., 2015. A Novel Morphological Model of Trabecular Bone Based on the Gyroid. Int J Appl Mech 07, 1550048. https://doi.org/10.1142/S1758825115500489

Rammohan, A.V., Tan, V.B.C., 2016. Morphological models of trabecular bone suitable for high-porosity regions and vertebrae. Comput Methods Biomech Biomed Engin 19, 1418–1422. https://doi.org/10.1080/10255842.2016.1146945

Reznikov, N., Chase, H., ben Zvi, Y., Tarle, V., Singer, M., Brumfeld, V., Shahar, R., Weiner, S., 2016. Inter-trabecular angle: A parameter of trabecular bone architecture in the human proximal femur that reveals underlying topological motifs. Acta Biomater 44, 65–72. https://doi.org/10.1016/j.actbio.2016.08.040

Roberts, A.P., Garboczi, E.J., 2002. Computation of the linear elastic properties of random porous materials with a wide variety of microstructure. Proceedings of the Royal Society A: Mathematical, Physical and Engineering Sciences 458, 1033–1054. https://doi.org/10.1098/rspa.2001.0900

Rodríguez-Montaño, Ó.L., Cortés-Rodríguez, C.J., Naddeo, F., Uva, A.E., Fiorentino, M., Naddeo, A., Cappetti, N., Gattullo, M., Monno, G., Boccaccio, A., 2019. Irregular Load Adapted Scaffold Optimization: A Computational Framework Based on Mechanobiological Criteria. ACS Biomater Sci Eng 5, 5392–5411. https://doi.org/10.1021/acsbiomaterials.9b01023

Ruiz, O., Schouwenaars, R., Ramírez, E.I., Jacobo, V.H., Ortiz, A., 2011. Effects of architecture, density and connectivity on the properties of trabecular bone: A two-dimensional, Voronoi cell based model study. AIP Conf Proc 1394, 77–89. https://doi.org/10.1063/1.3649938

Ruiz, O., Schouwenaars, R., Ramírez, E.I., Jacobo, V.H., Ortiz, A., 2010. Analysis of the architecture and mechanical properties of cancellous bone using 2D voronoi cell based models. WCE 2010 - World Congress on Engineering 2010 1, 609–614.

Saha, P.K., Xu, Y., Duan, H., Heiner, A., Liang, G., 2010. Volumetric topological analysis: A novel approach for trabecular bone classification on the continuum between plates and rods. IEEE Trans Med Imaging 29, 1821–1838. https://doi.org/10.1109/TMI.2010.2050779

Schaffner, G., Guo, X.D.E., Silva, M.J., Gibson, L.J., 2000. Modelling fatigue damage accumulation in

two-dimensional Voronoi honeycombs. Int J Mech Sci 42, 645–656. https://doi.org/10.1016/S0020-7403(99)00031-4

Si, H., 2015. TetGen, a Delaunay-Based Quality Tetrahedral Mesh Generator. ACM Trans. Math. Softw. 41. https://doi.org/10.1145/2629697

Silva, M.J., Gibson, L.J., 1997. Modeling the mechanical behavior of vertebral trabecular bone: Effects of age-related changes in microstructure. Bone 21, 191–199. https://doi.org/10.1016/S8756-3282(97)00100-2

Society, T.R., Transactions, P., Sciences, E., 1996. Triply periodic level surfaces as models for cubic tricontinuous block copolymer morphologies. Philosophical Transactions of the Royal Society A: Mathematical, Physical and Engineering Sciences 354, 2009–2023. https://doi.org/doi:10.1098/rsta.1996.0089

Soyarslan, C., Bargmann, S., Pradas, M., Weissmüller, J., 2018. 3D stochastic bicontinuous microstructures: Generation, topology and elasticity. Acta Mater 149, 326–340. https://doi.org/10.1016/j.actamat.2018.01.005

Stauber, M., Müller, R., 2006. Volumetric spatial decomposition of trabecular bone into rods and plates - A new method for local bone morphometry. Bone 38, 475–484. https://doi.org/10.1016/j.bone.2005.09.019

Stauber, M., Nazarian, A., Müller, R., 2014. Limitations of global morphometry in predicting trabecular bone failure. Journal of Bone and Mineral Research 29, 134–141. https://doi.org/10.1002/jbmr.2006

Timercan, A., Sheremetyev, V., Brailovski, V., 2021. Mechanical properties and fluid permeability of gyroid and diamond lattice structures for intervertebral devices: functional requirements and comparative analysis. Sci Technol Adv Mater 22, 285–300. https://doi.org/10.1080/14686996.2021.1907222

Vajjala, S., Kraynik, A.M., Gibson, L.J., 2000. A cellular solid model for modulus reduction due to resorption of trabeculae in bone. J Biomech Eng 122, 511–515. https://doi.org/10.1115/1.1289996

Wang, J., Zhou, B., Liu, X.S., Fields, A.J., Sanyal, A., Shi, X., Adams, M., Keaveny, T.M., Guo, X.E., 2015. Trabecular plates and rods determine elastic modulus and yield strength of human trabecular bone. Bone 72, 71–80. https://doi.org/10.1016/j.bone.2014.11.006

Yánez, A., Herrera, A., Martel, O., Monopoli, D., Afonso, H., 2016. Compressive behaviour of gyroid

lattice structures for human cancellous bone implant applications. Mater Sci Eng C Mater Biol Appl 68, 445–448. https://doi.org/10.1016/j.msec.2016.06.016

Yu, P., Du, Q., 2006. A VARIATIONAL CONSTRUCTION OF ANISOTROPIC MOBILITY IN PHASE-FIELD SIMULATION. DYNAMICAL SYSTEMS-SERIES B 6, 391–406.

Zheng, L., Kumar, S., Kochmann, D.M., 2021. Data-driven topology optimization of spinodoid metamaterials with seamlessly tunable anisotropy. Comput Methods Appl Mech Eng 383, 113894. https://doi.org/10.1016/J.CMA.2021.113894